# Classification of Breast Cancer Lesions in Ultrasound Images by using Attention Layer and loss Ensembles in Deep Convolutional Neural Networks

Elham Yousef Kalafi[1], Ata Jodeiri[2], Seyed Kamaledin Setarehdan[2], Ng Wei Lin[3], Kartini Binti Rahman[3], Nur Aishah Taib[4], Sarinder Kaur Dhillon[1]*

[1] Data Science & Bioinformatics Laboratory, Institute of Biological Sciences, Faculty of Science, University of Malaya, Kuala Lumpur, Malaysia

[2] School of Electrical and Computer Engineering, College of Engineering, University of Tehran, Tehran, Iran

[3] Department of Biomedical Imaging, Faculty of Medicine, University of Malaya, 50603 Kuala Lumpur, Malaysia

[4] Department of Surgery, Faculty of Medicine, University of Malaya, Kuala Lumpur, Malaysia

* Corresponding author: elham@um.edu.my, Data Science & Bioinformatics Laboratory, Institute of Biological Sciences, Faculty of Science, University of Malaya, Kuala Lumpur, Malaysia

## Abstract

Reliable classification of benign and malignant lesions in breast ultrasound images can provide an effective and relatively low cost method for early diagnosis of breast cancer. The accuracy of the diagnosis is however highly dependent on the quality of the ultrasound systems and the experience of the users (radiologists). The leverage in deep convolutional neural network approaches provided solutions in efficient analysis of breast ultrasound images.

In this study, we proposed a new framework for classification of breast cancer lesions by use of an attention module in modified VGG16 architecture. We also proposed new ensembled loss function which is the combination of binary cross-entropy and logarithm of the hyperbolic cosine loss to improve the model discrepancy between classified lesions and its labels. Networks trained from pretrained ImageNet weights, and subsequently fine-tuned with ultrasound datasets. The proposed model in this study outperformed other modified VGG16 architectures with the accuracy of 93% and also the results are competitive with other state of the art frameworks for classification of breast cancer lesions. In this study, we employed transfer learning approaches with the pre-trained VGG16 architecture. Different CNN models for classification task were trained to predict benign or malignant lesions in breast ultrasound images. Our Experimental results show that the choice of loss function is highly important in classification task and by adding an attention block we could empower the performance our model.

## Introduction

Breast cancer is the second leading cause of cancer death in women [1, 2]. Different types of imaging modalities such as mammography, ultrasound and magnetic resonance imaging have been used for diagnosing breast tumours. Whilst mammography has been proven to be a useful technique for diagnosing breast cancer leading to a reduced mortality [3], its sensitivity is limited in dense breast tissues. Breast density has been established as an independent risk of breast cancer [4–6]. Women with heterogenous dense and extremely dense breast tissues have relatively higher risks of 1.2 and 2.1 times in developing breast cancers compared to average women [7]. The accuracy rate of simple benign cysts diagnosis in breast ultrasound images has been reported to be 96–100%, so they do not require further evaluation [8]. In a meta-analysis of 29 studies, various adjunct screening methods have been studied to assess the limitation of various breast cancer screening modalities and ultrasound has demonstrated an increase of cancer detection by 40% [9].

Computer-aided diagnosis (CAD) systems are extensively used for detection and classification of tumours in breast ultrasound images. Statistical methods [10] have been predominantly used to analyse extracted features from lesion shape, margin, homogeneity and posterior acoustic attenuation. However, identification of shape and margin of lesions is difficult in ultrasound images [11]. Machine learning techniques have also been extensively deployed to analyse and classify lesions based on the handcrafted features consist of morphological and texture features of tumours [12, 13]. However, the extraction of features was still highly dependent on radiologist's experience. The struggles of researchers for handcrafting features has led to development of newer algorithms that can learn features automatically from data such as deep learning algorithms which are particularly strong tools for extracting non-linear features from data. Deep learning models are surprisingly promising in classification of ultrasound images, in which pattern recognition is not easily hand-engineered [14].

According to the rapid growth in deep learning based methods in last few years, one step further to efficiently integrate local and global features and exploit localised information [15] was employing attention mechanisms. The attention has been used in computer vision tasks such as detection [16, 17], segmentation [18] and classification [19], and it improves the model performance by focusing on the most relevant features that are important for the given task. To the best of our knowledge, attention modules have been widely used in medical image segmentation but not classification. In this study we used the attention gate module [15] in modified VGG16 architecture with new loss function to increase the classification performance for ultrasound breast lesion classification.

## Methods

The proposed framework in this study is inspired by [15], where the authors have introduced attention gates. We used one layer of attention in modified VGG16 architecture where the attention mechanism was applied to layer 13 just before pooling and the max pooling layer 18 as a last convolutional layer in the VGG16 feature extraction module [20]. The framework is illustrated in figure 1.

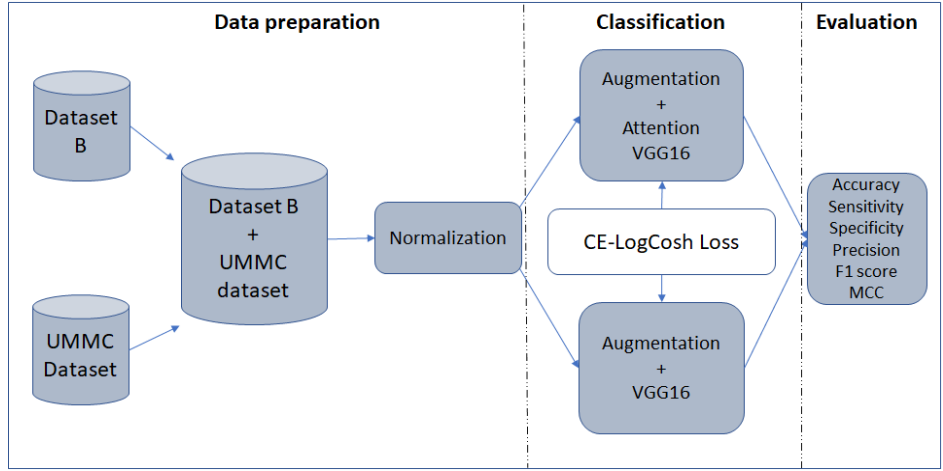

**Figure 1.** The overall framework of this study.

*The Dataset*

In this study we combined two datasets, one public breast ultrasound images, called as Database B with 163 images of 109 benign and 54 malignant lesions and the other dataset was collected at the University Malaya Medical Centre (UMMC), between June 2012 and April 2013 with 276 ultrasound breast images comprising 140 benign and 136 malignant lesions obtained from 83 different patients. All subjects were biopsy confirmed. The patients were either from the breast assessment clinic with palpable lumps or had sonographically detected lesions. Patients without confirmed histological diagnosis and those with a previously known histological diagnosis were excluded.

Most of the malignant lesions were infiltrating ductal carcinomas (IDCs), whereas the majority of the benign lesions were fibroadenomas. The sizes of the malignant lesions ranged from 0.5 to 9.0 cm (mean±SD: 2.1±1.2 cm), whereas the sizes of the benign lesions ranged from 0.3 to 5.0 cm (mean±SD: 1.4±1.0 cm).

*Pre-processing*

All ultrasound images were acquired using the Aixplorer ultrasound system (SuperSonic Imagine, Aix en Provence, France) using a 15-4 MHz linear transducer probe. Two specialized radiologists in breast imaging performed the scanning task and they were blinded to the histological diagnosis results. All images in UMMC dataset were in JPEG format and in the resolution of 1400x1050 pixels. The average image size in dataset B was 760 x 570 pixels where each of the images presented one or more lesions. In our experiments, the images were resampled to 128 x 128 pixels with a 75-15-10 train-test-validation split. Image normalization was applied to all of the images in the dataszsets to create a consistent dynamic range across the dataset. Figure 2, Illustrates the samples of benign and malignant lesions in breast ultrasound images.

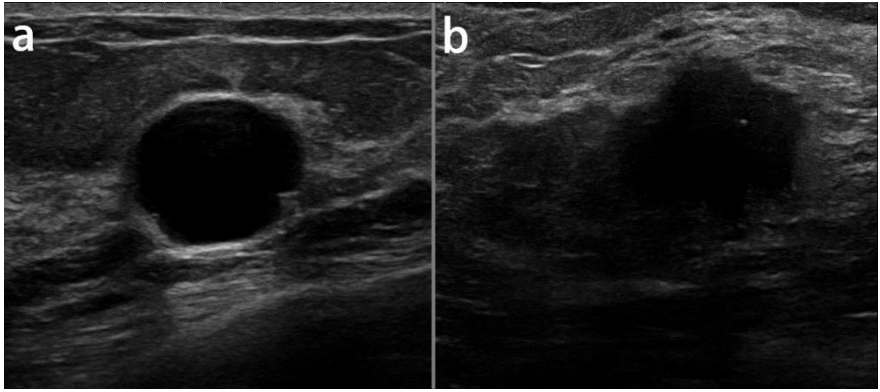

**Figure 2.** The samples of benign (a) and malignant (b) lesions in breast ultrasound images.

*Attention Module*

At the deep levels of convolutional layers, the network acquires the richest possible feature representation. Yet, spatial information may get lost in the high-level output maps with cascaded convolutions [21] or dense predictions are made on particular region of interest (ROI) and this approach leads to redundancy of low-level features extracted by all models within the cascade [22]. We used soft attention gate (AG) to deal with these issues. Through the AGs, the input feature map has element-wise multiplication with the attention coefficient to highlight the salient features (figure 3).

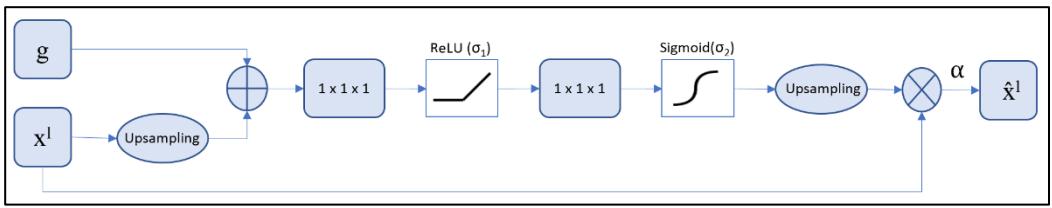

**Figure 3.** Illustration of attention gate (AG) adapted from [15]. Feature map upsampling is computed by bilinear interpolation.

The attention coefficients $\alpha_i \in [0, 1]$ is produced by AG at each pixel i to scale the input features $x_i^l$ and output features $\hat{x}_i^l$ at layer l. The localization of focus regions is obtained by the gating signal, g, for each pixel i. The gating signal is retrieved from coarser scale than the input features $x_i^l$. The linear attention coefficients are computed by the element-wise sum of $W_x$, $b_x$, $W_g$ and $b_g$ parameters followed by 1x1 linear transformation:

$$q_{attn}^l = \psi^T (\sigma_1 (W_x^T x_i^l + W_g^T g_i + b_g)) + b_\psi \quad (1)$$

$$\alpha_i^l = \sigma_2 (q_{attn}^l (x_i^l, g_i)) \quad (2)$$

ReLU and sigmoid as $\sigma_1$ and $\sigma_2$, respectively, are used to transform the intermediate maps in calculating the attention coefficients. The attention coefficients determine the important regions of image and prune features to maintain the relevant activations in specific task. The output maps at each scale are upsampled and then concatenated with the pruned features. In this stage, a 1x1x1 convolutions and non-linear activations are applied on each output map and then the high dimensional feature representation is supervised with CE-logcosh loss.

*Cross entropy - Log hyperbolic cosine (CE-LogCosh) Loss*

According to importance of the loss function in learning algorithm, towards having better learning system, this study is inspired by the ensembled methods [23] in order to develop an ensemble loss function. We combined two loss functions, cross entropy [24] and log hyperbolic cosine [25] to boost the learning process and achieving better performance. The cross entropy loss, compares the distribution of predictions and true labels and defines as:

$$L_{CE}(y, \hat{y}) = - \sum_i y_i \log(\hat{y}_i) \quad (3)$$

The log-cosh loss function is the hyperbolic cosine algorithm of the prediction error.

$$L_{LCH}(y, \hat{y}) = \sum_i \log(\cos h (\hat{y}_i - y_i)) \quad (4)$$

Where y is the label and $\hat{y}_i$ is the predicted label. The proposed ensembled loss function is as:

$$\text{CE-logcosh Loss} = \alpha L_{CE} + \beta L_{LCH} \quad (5)$$

In CE-logcosh Loss function, α and β are parameters that can tuned to shift the emphasis on cross entropy or logcosh loss. In this study we set α and β to 0.5 as the best performance was achieved.

*Network Architecture*

In this study, we used convolutional layers in VGG16 to extract features from the datasets. Figure 4 is the schematic of proposed network architecture in which pre-trained VGG16 were used for fine tuning and feature extraction. The feature maps in layer 13 and 18 were then used in the attention block and then the output was fed to modified fully connected (FC) [26] layers for classification of malignant and benign lesions.

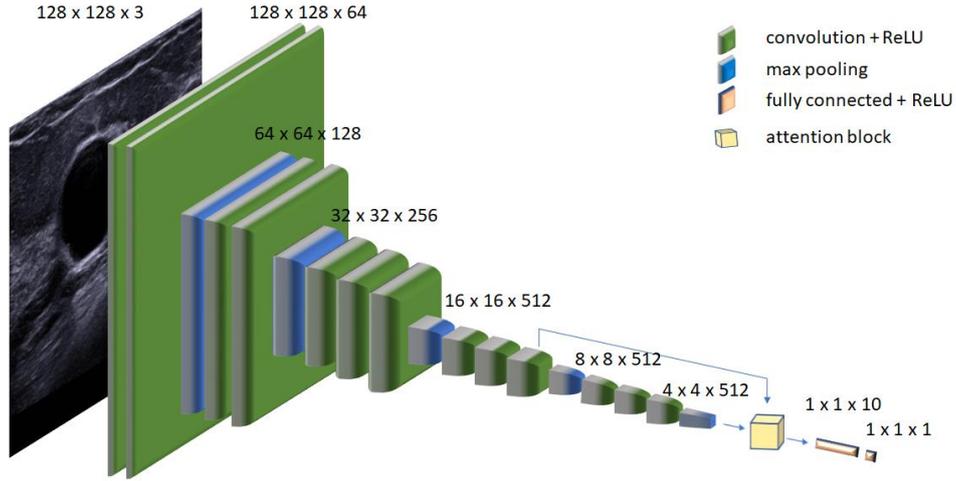

**Figure 4.** Proposed attention-VGG16 with attention block.

The number of breast ultrasound images in this study is relatively low, hence establishing deeper networks and training networks from scratch is not feasible. ImageNet dataset was used to pre-train VGG16 and we replaced the dense layers and the last 1000 – way fully connected layer with a new FC layers, using ImageNet pretrained weights W drawn from a normal distribution as follows: $W \sim N(\mu = 1; \sigma^2 = 0.01)$. The experiment was trained for 250 epochs with a batch size of 32. The model was optimized using RMSprop with initial learning rate of $2 \times 10^{-6}$ which decays by $10^{-6}$ on every epoch. All experiments are programmed using the Keras framework with TensorFlow backend.

We proposed new model based on attention gating and new loss function to enhance the performance of classification for breast ultrasound images. The "dropout" strategy [27] was also used to avoid overfitting.

*Evaluation*

Classification performance of models in this study were measured by sensitivity, specificity, accuracy, precision, F1 score and Matthews Correlation Coefficient [28], which were obtained from confusion matrix entries. In a confusion matrix, the relation between classification outcomes and predicted classes are illustrated. The level of classification performance is calculated by the number of correct and incorrect classified samples in each class. Accuracy is computed based on the total number of correct predictions, defined as:

$$\frac{TP+TN}{TP+FN+TN+FP} \qquad (6)$$

Sensitivity is the proportion of true positive that are identified correctly, defined as:

$$\frac{TP}{TP+FN} \tag{7}$$

Specificity is the proportion of true negative that are correctly predicted, defined as:

$$\frac{TN}{TN+FP} \tag{8}$$

Precision or positive predictive value, is the ratio of correctly predicted positive observations to total predicted positive observations, defined as:

$$\frac{TP}{TP+FP} \tag{9}$$

F1 score is the weighted average of precision which is calculated as:

$$\frac{2TP}{2TP+FP+FN} \tag{10}$$

Matthews Correlation Coefficient (MCC) is correlation coefficient between the observed and predicted classifications, defined as:

$$\frac{TP*TN - FP*FN}{\sqrt{(TP+FP)*(TP+FN)*(TN+FP)*(TN+FN)}} \tag{11}$$

Where True Positive (TP) and True Negative (TN) stand for the number of correct predictions and False Positive (FP) and False Negative (FN) that of incorrect predictions.

## Results

We evaluated our proposed model on classification of ultrasound breast lesions to benign and malignant. In particular, correct classification of benign and malignant lesions is difficult task because of variety in shape and poor contrast on ultrasound breast images. We compared our model and standard VGG16 with different losses in terms of classification performance.

From table 1 and figure 5, it is notable that our proposed model with CE-logcosh outperformed other classification models in terms of accuracy, sensitivity, specificity, precision, F1 score and MCC.

**Table 1.** The comparison of the VGG16 and attention-VGG16 models with different loss functions in classification of benign and malignant lesions.

| Models | Loss | Sensitivity | Specificity | Precision | Accuracy | F1 Score | MCC |
|---|---|---|---|---|---|---|---|
| VGG16 | CE | 0.80 | 0.78 | 0.84 | 0.80 | 0.82 | 0.59 |
| VGG16 | Logcosh | 0.84 | 0.80 | 0.84 | 0.82 | 0.84 | 0.62 |
| VGG16 | CE-logcosh | 0.95 | 0.82 | 0.84 | 0.89 | 0.89 | 0.79 |
| Attention-VGG16 | CE | 0.88 | 0.85 | 0.88 | 0.87 | 0.88 | 0.73 |
| Attention-VGG16 | Logcosh | 0.85 | 0.84 | 0.88 | 0.84 | 0.86 | 0.68 |
| Attention-VGG16 | CE-logcosh | **0.96** | **0.90** | **0.92** | **0.93** | **0.94** | **0.87** |

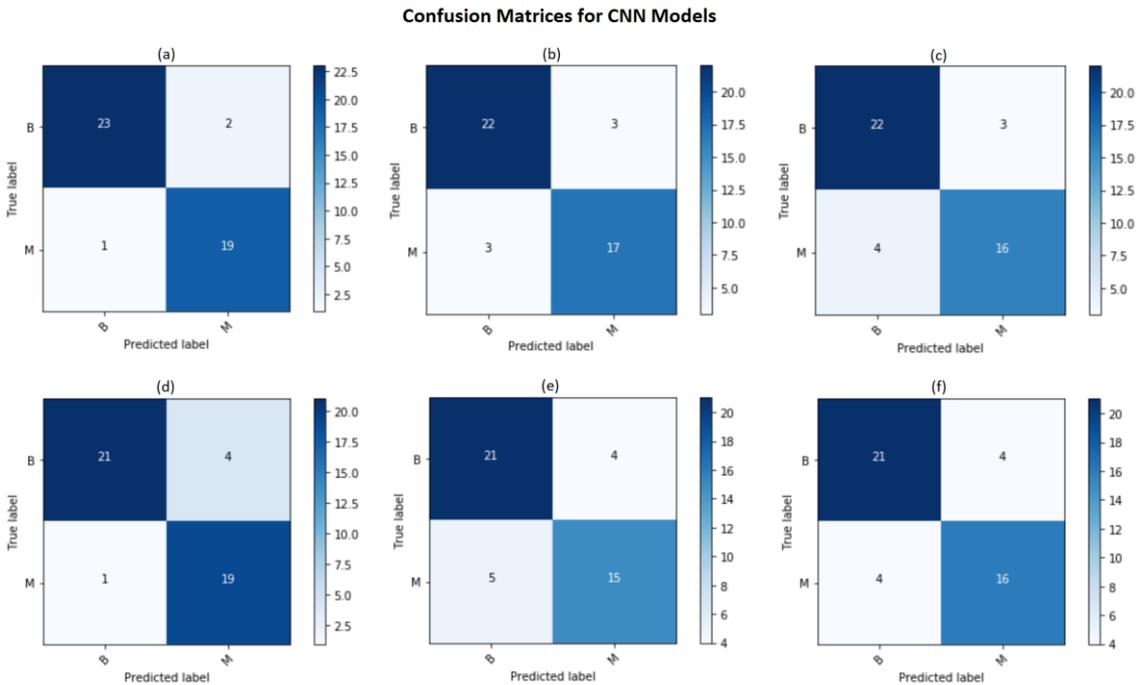

**Figure 5.** Confusion matrices for different models. (a) attention VGG16 with CE-logcosh loss, (b) attention VGG16 with CE loss, (c) attention VGG16 with logcosh loss, (d) VGG16 with CE-logcosh loss (e) VGG16 with CE loss, (f) VGG16 with logcosh loss.

## Discussion

In this paper, modified VGG16 architectures were compared in order to achieve higher performance in classification of benign and malignant breast tumours. Modifications such as additional attention block, different dense layers and ensembled loss functions were made. One of the improvements in the CNN models was the use of ensembled loss functions. Within the training phase, in the gradient propagation optimization, the weight of each loss function was tuned and they were parametrized by α and β to control the emphasis. To the best of our knowledge, logcosh loss works mostly like $L_2$ at small values and like $L_1$ at large values and is usually used in regression or reconstruction tasks [25]. In this study we used logcosh loss, combined with binary cross-entropy to improve the classification accuracy. As it is notable in table 1, the ensemble of both losses could improve the performance of classification.

On the other hand, by using attention block, relevant spatial information is identified from low-level feature maps and propagated to classification stage. The lack of these relevant spatial information is caused by transforming the large size of the feature maps that are obtained after the convolutional layers in the CNN and reaching smaller feature dimensions. Therefore, the attention block was proposed which attempts to compute the contribution of each feature.

In our study, out of all the models, the attention VGG16 with logcosh loss has demonstrated the highest accuracy and precision. Additionally, the proposed deep convolutional neural network architecture does not need prior expert knowledge or image segmentation, hence it will be more convenient in CAD and suitable for future clinical diagnosis.

Table 2 demonstrate some state-of-the-art deep learning models in lesion classification for breast ultrasound images. It is notable that the performance of our proposed model is comparable to these models.

**Table 2.** The state of the art of deep learning models in breast ultrasound lesion classification.

| References | Dataset | Deep learning Models | Performance |
|---|---|---|---|
| [29] | 4254 benign 3154 malignant | GoogLeNet | Accuracy: 91.23% Sensitivity: 84.29% Specificity: 96.07% |
| [30] | 135 benign 92 malignant | Boltzmann | Accuracy: 93.4% Sensitivity: 88.6% Specificity: 97.1% |
| [31] | 100 benign 100 malignant | Deep Polynomial network+SVM | Accuracy: 92.40% Sensitivity: 92.67% Specificity: 91.36% |
| [32] | 275 benign 245 malignant | Stacked denoising Autoencoder | Accuracy: 82.4% Sensitivity: 78.7% Specificity: 85.7% |
| Current Study | 249 benign 190 malignant | Attention VGG16 + ensembled loss | Accuracy: 93% Sensitivity: 96% Specificity: 90% |

One of the hyperparameters that was assessed in this study was the number of neurons in the dense layers. We used the smallest number of neurons to decrease the number of parameters and surprisingly this achieved same accuracy while using 4096 or 10 neurons in dense layers.

In summary, we proposed the attention VGG16 classifier as a potential architecture in classifying breast cancer ultrasound images. Having said this, we suggest that this model is tested further using a larger dataset to improve the robustness of this architecture. Additionally, we also suggest that the VGG16 to be implemented with machine learning classifiers as potential architectures in clinical studies. In future studies, the deep convolutional neural networks architecture should be conducted on a larger image data with various tumor subtypes to adapt it to multi-class classification as the classification of breast lesions` subtypes is of greater clinical impact [33, 34].

## Conclusion

In this study, we analysed some computer-aided diagnosis models for classification of benign and malignant lesions on UMMC breast ultrasound image dataset. We employed transfer learning approaches with the pre-trained VGG16 architecture. Different CNN models for classification task were trained to predict benign or malignant lesions in breast ultrasound images. Our Experimental results demonstrated that the choice of loss function is highly important in classification task and by adding an attention block we could empower the performance our model. Our proposed model with extracted features from VGG16 and fully connected network with only 10 neurons achieved the best performance in classification task with respect to the precision of 92% and accuracy of 93%. With this framework, evaluation tests show that the combination of loss functions can provide suitable information to enable the construction of the most accurate prediction model when compared with other models. In the future, other deep neural network models will be tested on a larger dataset of ultrasound images with the hope to further increase the accuracy of performance.

## References

1. Spaeth E, Starlard-Davenport A, Allman R (2018) Bridging the Data Gap in Breast Cancer Risk Assessment to Enable Widespread Clinical Implementation across the Multiethnic Landscape of the US. J Cancer Treat Diagn 2:1–6. https://doi.org/10.29245/2578-2967/2018/4.1137

2. Joshi R, Basu A (2018) The Involvement of S6 Kinase-2 in Breast Cancer. Res Apprec Day

3. Tabár L, Fagerberg CJ, Gad A, et al (1985) Reduction in mortality from breast cancer after mass screening with mammography. Randomised trial from the Breast Cancer Screening Working Group of the Swedish National Board of Health and Welfare. Lancet Lond Engl 1:829–832

4. Houssami N, Abraham LA, Kerlikowske K, et al (2013) Risk factors for second screen-detected or interval breast cancers in women with a personal history of breast cancer participating in mammography screening. Cancer Epidemiol Biomark Prev Publ Am Assoc Cancer Res Cosponsored Am Soc Prev Oncol 22:946–961. https://doi.org/10.1158/1055-9965.EPI-12-1208-T

5. Kerlikowske K, Zhu W, Tosteson ANA, et al (2015) Identifying women with dense breasts at high risk for interval cancer: a cohort study. Ann Intern Med 162:673–681. https://doi.org/10.7326/M14-1465

6. McCormack VA, dos Santos Silva I (2006) Breast density and parenchymal patterns as markers of breast cancer risk: a meta-analysis. Cancer Epidemiol Biomark Prev Publ Am Assoc Cancer Res Cosponsored Am Soc Prev Oncol 15:1159–1169. https://doi.org/10.1158/1055-9965.EPI-06-0034

7. Sickles EA (2010) The use of breast imaging to screen women at high risk for cancer. Radiol Clin North Am 48:859–878. https://doi.org/10.1016/j.rcl.2010.06.012

8. Jackson VP (1990) The role of US in breast imaging. Radiology 177:305–311. https://doi.org/10.1148/radiology.177.2.2217759

9. Rebolj M, Assi V, Brentnall A, et al (2018) Addition of ultrasound to mammography in the case of dense breast tissue: systematic review and meta-analysis. Br J Cancer 118:1559–1570. https://doi.org/10.1038/s41416-018-0080-3

10. Giger ML, Al-Hallaq H, Huo Z, et al (1999) Computerized analysis of lesions in US images of the breast. Acad Radiol 6:665–674. https://doi.org/10.1016/S1076-6332(99)80115-9

11. Sivaramakrishna R, Powell KA, Lieber ML, et al (2002) Texture analysis of lesions in breast ultrasound images. Comput Med Imaging Graph 26:303–307. https://doi.org/10.1016/S0895-6111(02)00027-7

12. Xiao Y, Zeng J, Niu L, et al (2014) Computer-aided diagnosis based on quantitative elastographic features with supersonic shear wave imaging. Ultrasound Med Biol 40:275–286. https://doi.org/10.1016/j.ultrasmedbio.2013.09.032

13. Zhang Q, Xiao Y, Chen S, et al (2015) Quantification of elastic heterogeneity using contourlet-based texture analysis in shear-wave elastography for breast tumor classification. Ultrasound Med Biol 41:588–600. https://doi.org/10.1016/j.ultrasmedbio.2014.09.003

14. Brattain LJ, Telfer BA, Dhyani M, et al (2018) Machine learning for medical ultrasound: status, methods, and future opportunities. Abdom Radiol 43:786–799. https://doi.org/10.1007/s00261-018-1517-0

15. Schlemper J, Oktay O, Schaap M, et al (2019) Attention gated networks: Learning to leverage salient regions in medical images. Med Image Anal 53:197–207. https://doi.org/10.1016/j.media.2019.01.012

16. Cao C, Liu X, Yang Y, et al (2015) Look and Think Twice: Capturing Top-Down Visual Attention with Feedback Convolutional Neural Networks. In: 2015 IEEE International Conference on Computer Vision (ICCV). pp 2956–2964

17. Zhou B, Khosla A, Lapedriza A, et al (2015) Learning Deep Features for Discriminative Localization. ArXiv151204150 Cs

18. Simonyan K, Vedaldi A, Zisserman A (2014) Deep Inside Convolutional Networks: Visualising Image Classification Models and Saliency Maps. ArXiv13126034 Cs

19. Zagoruyko S, Komodakis N (2017) Paying More Attention to Attention: Improving the Performance of Convolutional Neural Networks via Attention Transfer. ArXiv161203928 Cs

20. Simonyan K, Zisserman A (2015) Very Deep Convolutional Networks for Large-Scale Image Recognition. ArXiv14091556 Cs

21. Abraham N, Khan NM (2018) A Novel Focal Tversky loss function with improved Attention U-Net for lesion segmentation. ArXiv181007842 Cs

22. Oktay O, Schlemper J, Folgoc LL, et al (2018) Attention U-Net: Learning Where to Look for the Pancreas

23. Hajiabadi H, Molla-Aliod D, Monsefi R (2017) On Extending Neural Networks with Loss Ensembles for Text Classification. In: Proceedings of the Australasian Language Technology Association Workshop 2017. Brisbane, Australia, pp 98–102

24. Murphy KP, Bach F (2012) Machine Learning: A Probabilistic Perspective. MIT Press, Cambridge, MA

25. Chen P, Chen G, Zhang S (2018) Log Hyperbolic Cosine Loss Improves Variational Auto-Encoder

26. Krizhevsky A, Sutskever I, Hinton GE (2017) ImageNet Classification with Deep Convolutional Neural Networks. Commun ACM 60:84–90. https://doi.org/10.1145/3065386

27. Hinton GE, Srivastava N, Krizhevsky A, et al (2012) Improving neural networks by preventing co-adaptation of feature detectors. ArXiv12070580 Cs

28. Powers DM (2011) Evaluation: from Precision, Recall and F-measure to ROC, Informedness, Markedness and Correlation

29. Han S, Kang H-K, Jeong J-Y, et al (2017) A deep learning framework for supporting the classification of breast lesions in ultrasound images. Phys Med Biol 62:7714–7728. https://doi.org/10.1088/1361-6560/aa82ec

30. Zhang Q, Xiao Y, Dai W, et al (2016) Deep learning based classification of breast tumors with shear-wave elastography. Ultrasonics 72:150–157. https://doi.org/10.1016/j.ultras.2016.08.004

31. Shi J, Zhou S, Liu X, et al (2016) Stacked deep polynomial network based representation learning for tumor classification with small ultrasound image dataset. Neurocomputing 194:87–94. https://doi.org/10.1016/j.neucom.2016.01.074

32. Cheng J-Z, Ni D, Chou Y-H, et al (2016) Computer-Aided Diagnosis with Deep Learning Architecture: Applications to Breast Lesions in US Images and Pulmonary Nodules in CT Scans. Sci Rep 6:24454. https://doi.org/10.1038/srep24454

33. Zhang Q, Xiao Y, Chen S, et al (2015) Quantification of elastic heterogeneity using contourlet-based texture analysis in shear-wave elastography for breast tumor classification. Ultrasound Med Biol 41:588–600. https://doi.org/10.1016/j.ultrasmedbio.2014.09.003

34. Berg WA, Mendelson EB, Cosgrove DO, et al (2015) Quantitative Maximum Shear-Wave Stiffness of Breast Masses as a Predictor of Histopathologic Severity. AJR Am J Roentgenol 205:448–455. https://doi.org/10.2214/AJR.14.13448